\documentstyle[emulateapj,newapa]{article} 
\catcode`\@=11 
\def\lsim{\mathrel{\mathpalette\@versim<}}
\def\gsim{\mathrel{\mathpalette\@versim>}}
\def\be{\begin{equation}}
\def\ee{\end{equation}}
\def\@versim#1#2{\vcenter{\offinterlineskip
        \ialign{$\m@th#1\hfil##\hfil$\crcr#2\crcr\sim\crcr } }}
\catcode`\@=12 

\eqsecnum

\begin{document}

\date{}  
\submitted{Accepted for publication in Astrophysical Journal Letters}
\title{Dust Echos from Gamma Ray Bursts}
\author{Ann A. Esin\altaffilmark{1} and Roger Blandford}
\affil{130-33 Caltech, Pasadena, CA 91125; aidle@tapir.caltech.edu, 
rdb@tapir.caltech.edu}
\altaffiltext{1}{Chandra Fellow}

\begin{abstract} 
The deviation from the power-law decline of the optical flux observed
in GRB~970228 and GRB~980326 has been used recently to argue in favor of the
connection between GRBs and supernovae.  We consider an alternative
explanation for this phenomenon, based on the scattering of a prompt
optical burst by $0.1M_{\odot}$ of dust located beyond its sublimation
radius $0.1-1\,{\rm pc}$ from the burst.  In both cases, the optical
energy observed at the time of the first detection of the afterglow
suffices to produce an echo after $\sim20-30$~d, as observed. Prompt
optical monitoring of future bursts and multiband photometry of the
afterglows will enable quantitative tests of simple models of dust
reprocessing and a prediction of the source redshift.
\end{abstract} 

\keywords{gamma rays: bursts -- ISM: dust, extinction}

\section{Introduction}
\label{intro}

The relationship between Gamma Ray Bursts (GRBs) and supernovae has
become increasingly interesting over the past year.  Though exploding
massive stars have long been considered as possible progenitors of
GRBs (e.g. \fcitep{woo93}), no evidence existed to support these
theories until observations of the afterglow of GRB~980425 suggested
an association of the burst with an unusual supernova 1998bw
(\fcitep{gea98}, \fcitep{kea98}).  Later reanalysis of the optical
afterglow lightcurves of two other bursts, GRB~970228 (\fcitep{fea99}) and
GRB~980326 (\fcitep{bea99}) showed a deviation from the power-law decline
expected if the emission is due to synchrotron radiation from
electrons accelerated by the blast wave.  In both cases a significant
excess emission was observed around $\sim 30$ days after the gamma-ray
burst, with simultaneous reddening of the spectrum. \fcitet{bea99},
\fcitet{gea99}, and \fcitet{rei99} attribute this excess to the
emission from an underlying supernova event.

The relationship of GRBs to SN explosions is a question of great
importance, since it provides a powerful clue to the fundamental
nature of these objects.  However, the evidence presented so far is
circumstantial -- the association of GRB~980425 with SN 1998bw is
unproven and the excess emission seen from GRB~970228 and GRB~980326 is based
upon relatively few actual measurements -- and possible alternative
explanations need to be seriously considered, if only to strengthen
the case for the SN explanation.  In this spirit, \fcitet{wad99}
suggested that the red excess emission observed in GRB~970228 and GRB~980326
is due to dust in the vicinity of the burst progenitor absorbing and
then re-radiating the optical/UV flash observed shortly after the
recent GRB~990123 (\fcitep{aea99}) and generally attributed to the
reverse shock which propagates into the fireball ejecta
(\fcitep{mrp94}, \fcitep{mer97}, \fcitep{pam98}, \fcitep{sap99}).
However, the \acite{wad99} scenario has two shortcomings.  Firstly,
the equilibrium temperature of dust is limited to $\sim2300$~K and so
the emission should peak at $\sim 2\,(1+z)\,{\rm \mu m}$ (where $z$ is
the GRB redshift), although a small amount of higher temperature
emission may be produced by the dust as it is subliming.  Secondly,
the optical flash is so powerful that the sublimation radius lies
beyond $\sim10$~pc from the GRB. Thus, in this picture it is rather
difficult to reproduce the observed flux in the $0.4-0.8\,{\rm \mu m}$
band with a time delay of order a few weeks.

In this letter we propose an alternative explanation, which relies on
the {\em scattering} of the direct optical transient emitted in the
first day by dust as the primary source of excess optical radiation.
The fundamental point is that in the two observed cases, assuming
isotropic emission, the fluence of the {\it observed transient}
exceeds that of the reported excess and the {\it unobserved transient}
is even larger if we extrapolate to earlier times. A fraction of this
emission scattered from a radius where dust can outlive the optical
transient should therefore produce a delayed echo. As dust absorbs
selectively as well as scatters, the echo is likely to be
significantly redder than the original optical transient, as reported.

In the next section, we describe our model for the dust scattering
properties and then present the results in the context of the observed
GRBs in \S\ref{obs}. Implications for future tests of our scenario are
discussed in \S\ref{impl}.  We assume $h=0.6$, $\Omega_{\rm M} = 0.3$,
and $\Omega_{\Lambda} = 0.7$ so that the angular diameter distance of
the GRB is $D_{\rm A}=1.5-2$~Gpc for $0.5\lsim z\lsim3$.
\section{Dust Echos}
\label{dust}
\subsection{Sublimation Radius}
\cite{wad99} estimate that dust grains in the path
of the optical/UV flash will be effectively sublimed out to a
distance 
\begin{equation}
\label{rsubl}
R_{\rm sub} \sim 1\,(Q_{\rm abs} L_{47}/a_{-1})^{1/2}\,{\rm pc}, 
\end{equation}
where $Q_{\rm abs} \sim 1$ is the absorption efficiency factor for
optical/UV photons, $L_{47}\equiv\int d\nu L_\nu/10^{47}\,{\rm erg\
s^{-1}}$ is the unbeamed luminosity of the optical transient (OT) in
the $1-7.5\,{\rm eV}$ energy band, $a_{-1}\sim1$ is the dust grain
size in units of $0.1\,{\rm \mu m}$. Beyond $R_{\rm sub}$, only the
most refractory grains, like silicates, can survive. Note that the
thermal time for a typical dust particle is of order
$10^{-4}-10^{-2}\,{\rm s}$, much shorter that the duration of the
optical transient, so that we can treat grains as being in thermal
equilibrium with the incident radiation, (which has a pressure $P_{\rm
sub}\sim0.03$~dyne cm$^{-2}$ at $R_{\rm sub}$).

The extinction properties of silicate dust particles were computed by
\fcitet{drl84} (see their Fig. 10), for a power-law distribution of
particle sizes proposed by \fcitet{mrn77} to explain interstellar
starlight extinction.  Based on their results, we take the ratio of
the scattering and absorption efficiency factors to be of order
$Q_{\rm sc}/Q_{\rm abs} \simeq 4$, and the average scattering angle to
be $\langle\cos{\theta}\rangle \equiv\langle\mu\rangle\simeq 0.5$ for
observed wavelengths $0.2-1\ (1+z)\,{\rm \mu m}$.
\includegraphics{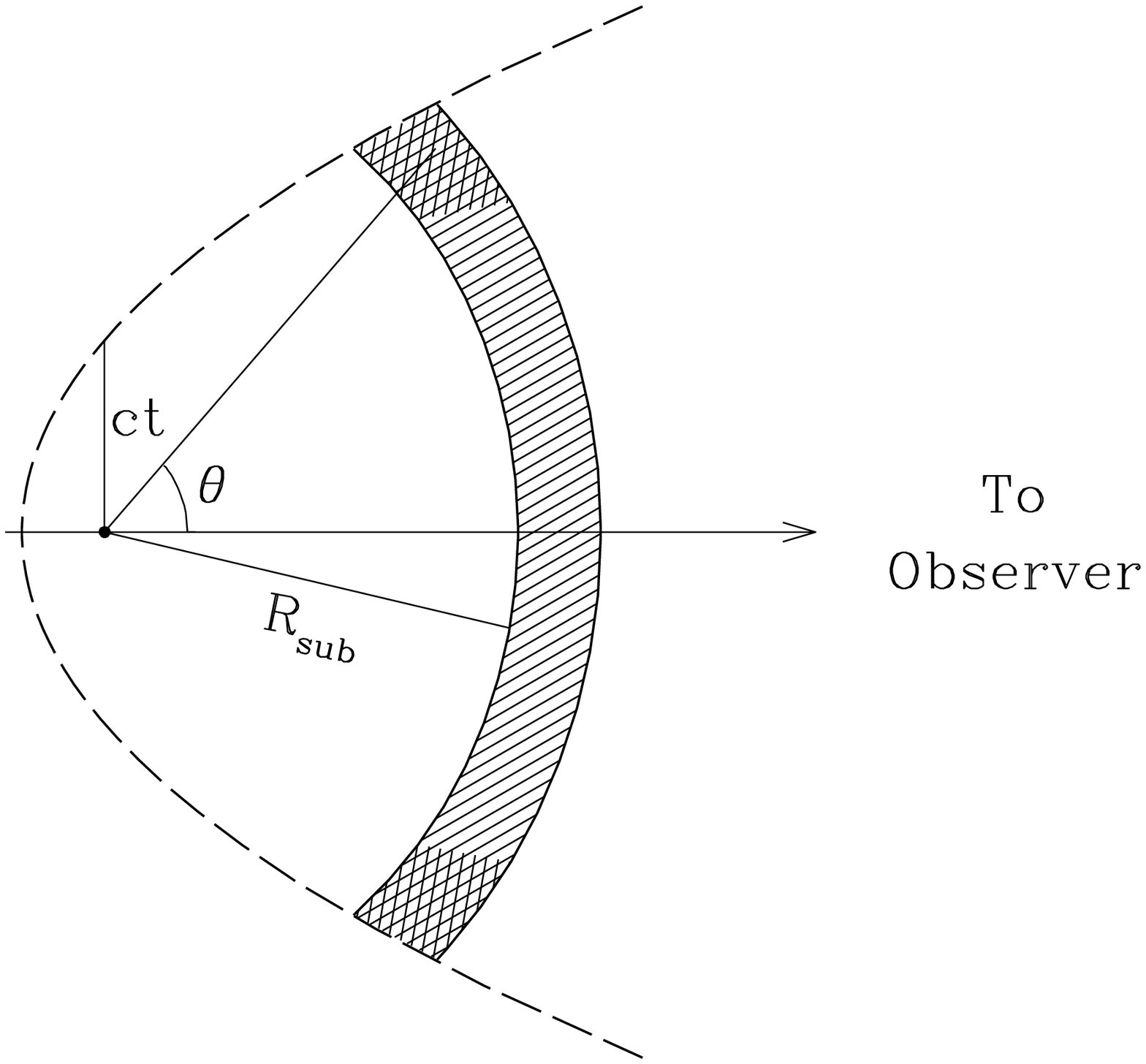}
\vskip 2.7in 
\figcaption{\label{fig1} A schematic diagram of the GRB environment.
The long-dashed line represents the position of the expanding
optical/UV photon front at time $t$ in the frame of the GRB.  The
shaded area shows the region where the dust is not sublimated
instantaneously.  The heavily shaded area shows the region of the
shell from which the scattered radiation is observed while the lightly
shaded area represents the regions where the dust is not scattering
any more.}
\vskip 0.2in
\subsection{Source Geometry}
Fig. \ref{fig1} shows a schematic picture of the GRB environment
observed at time $t$ after the detection of $\gamma$-rays.  The
incident optical transient emission is supposed to be limited to an
interval $\Delta t^{\rm OT}\equiv\Delta t^{\rm OT}_{\rm ob}/(1+z)<<t$
after the GRB, and scattered by dust beyond $R_{\rm sub}$.  We
specialize immediately to the case when the dust is associated with an
outflowing spherical wind, and the OT is isotropic.  (It is
straightforward to modify our formalism to accommodate other
reasonable assumptions, as discussed in \fcitet{mbr99}.) As the dust
density declines with distance as $R^{-2}$, the light ``echo''
observed at time $t_{\rm ob}=t(1+z)$ will be scattered by dust
concentrated in a ring located at the intersection of the sphere
$R=R_{\rm sub}$ and the paraboloid
\begin{equation}
\label{time}
t = \frac{R}{c} (1-\mu),
\end{equation}
where $\mu=\cos\theta$ (see Fig. \ref{fig1}). In our model, it is
adequate to ignore a finite re-processing time and the radial
distribution of the dust.  The dust only has to survive for a time
$\sim\Delta t^{\rm OT}$.  We expect that, in practice, it will be quickly
destroyed by the effects of secondary cosmic ray electrons created
through electron scattering of the GRB so that the observed optical
afterglow need not necessarily be subject to the same extinction as the
echo.
\subsection{Optical Scattering}
\label{scat}
The optical echo flux density $F_{\rm\nu_{ob}}^{\rm E}$,
observed at frequency $\nu_{\rm ob} = \nu (1+z)^{-1}$ is
\begin{eqnarray}
\label{Fsc}
F^{\rm E}_{\nu_{\rm ob}}(\nu_{\rm ob},t_{\rm ob}) 
= \frac{L_\nu(\nu,t)}{2(1+z)^3
D_{\rm A}^2} \frac{c\Delta t^{\rm OT}}{R} \frac{d {\mathcal{P}}^{\rm
sc}}{d {\Omega}}(\nu, \mu);\\ \nonumber
0<t_{\rm ob}<2R_{\rm sub}(1+z)/c,
\end{eqnarray}
where $d{\mathcal{P}}^{\rm sc}(\nu, \mu)/d\Omega$, is the 
probability of escape along the direction defined by angle 
$\theta = \cos^{-1}{\mu}$ for a photon of frequency $\nu$.

\includegraphics{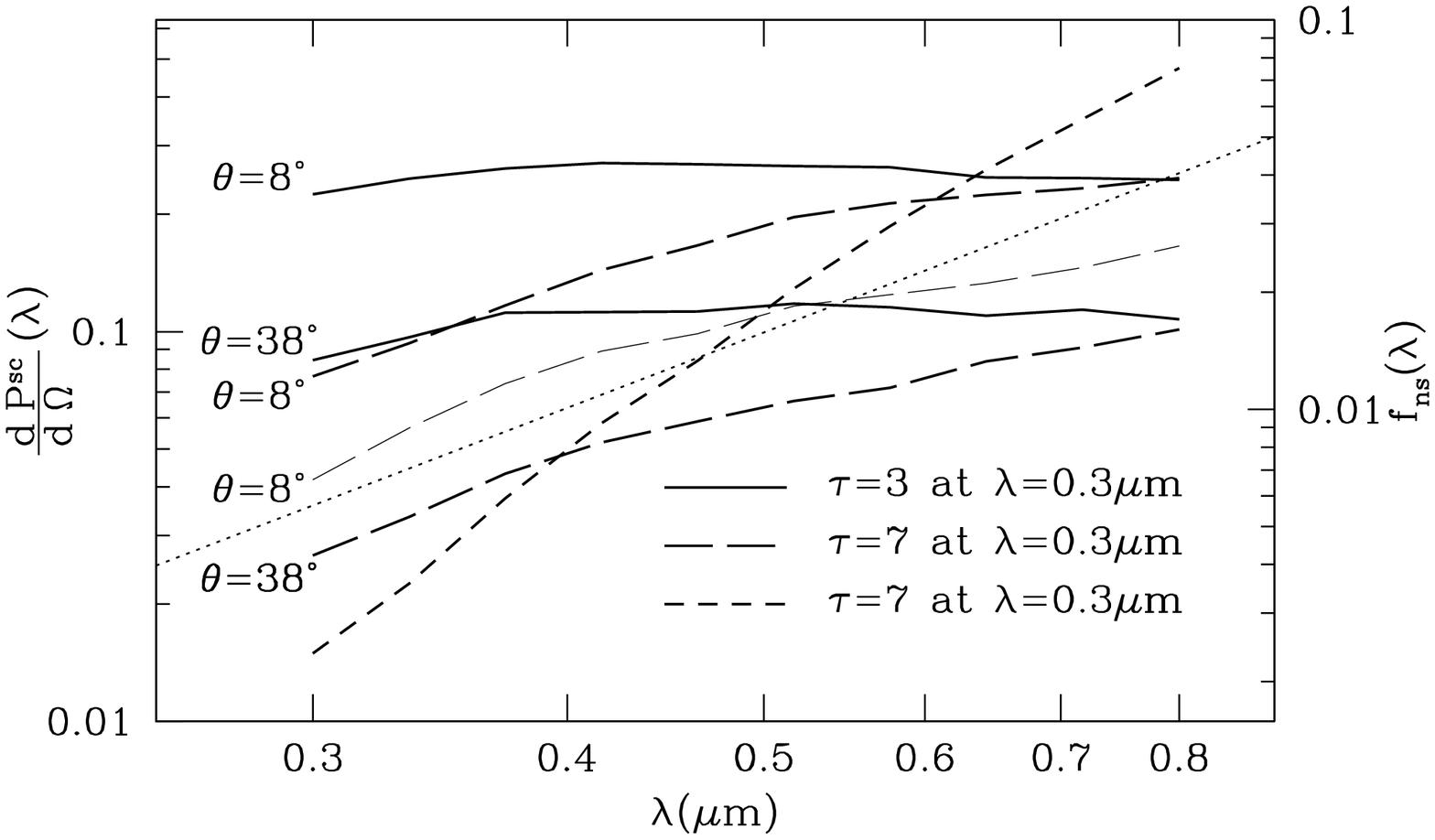} 
\vskip 2.2in 
\figcaption{\label{fig2} Solid and long-dashed lines show the escape
probability for photons scattered by a dust slab for different values
of $\theta$ and $\tau$ (as marked in the figure).  For comparison, the
dotted line represents an escape probability that increases the
spectral index of the echo relative to the OT by 2. The thin
long-dashed line shows the results computed using a different
$d\sigma/d\Omega \propto 1 + 2 \mu + \mu^2$ which gives similar
results although it is less peaked at $\mu\sim1$. The short-dashed
line shows the fraction of photons at each wavelength, $f_{\rm
ns}(\lambda)$ which pass through the slab unscattered, approximating
the escape probability at $\theta=0$.}
\vskip 0.2in

From Eq. (\ref{Fsc}) it is clear that the only time dependence 
comes from the angular dependence of the escape probability, 
$d{\mathcal{P}}^{\rm sc}/d {\Omega}$; $F^{\rm sc}_{\nu_{\rm ob}}
(t_{\rm ob})$ is simply a step function for isotropic scattering.
We adopt a Henyey-Greenstein function (e.g. \fcitep{whi79}) to describe the
differential cross section for dust scattering:
\begin{equation}
\label{sigma}
\frac{d \sigma}{d \Omega} \propto \frac{1-\langle\mu\rangle^2}
{[1+\langle\mu\rangle^2-2 \langle\mu\rangle \mu]^{3/2}},
\end{equation}
with $\langle \mu \rangle = 0.5$ (\fcitep{drl84}).  We then use 
Eq. (\ref{sigma}) to compute $d{\mathcal{P}}^{\rm sc} (\lambda,
\mu)/d\Omega$ numerically for a slab-like dust cloud.
The results for different observer
angles (w.r.t. to the slab normal vector) and two different values of
the total extinction, $\tau = \tau_{\rm abs} + \tau_{\rm sc}$ (measured at
$\lambda=0.3\,{\rm \mu m}$) are shown in Fig. \ref{fig2}.  The
differential escape probability is normalized so that the integral
$\int_{4 \pi} {\frac{d {\mathcal{P}}^{\rm sc}}{d {\Omega}}(\lambda,
\mu)\, d \Omega}$ is equal to the escape probability from the dust
cloud.  Fig. \ref{fig2} shows that at low optical depth $\tau_{0.3}
\equiv\tau(0.3\,\mu{\rm m})\lsim3$, the echo should have a similar
color to the OT whereas at larger optical depth, the echo will be much
redder due to absorption.

To illustrate that our calculation of the escape probability is not
overly simplified (though it ignores wavelength dependence of the
functional form for $d\sigma/d \Omega$,
(\fcitep{whi79}) in Fig. \ref{fig2} we show one curve (thin
long-dashed line) computed using $d\sigma/d\Omega \propto 1 + 2 \mu +
\mu^2$. This expression gives the same value of $\langle \mu \rangle$
but is less strongly peaked at $\mu=1$ than Eq. (\ref{sigma}).  The
resulting $d{\mathcal{P}}^{\rm sc} (\lambda, \mu)/d\Omega$ is very similar 
to what we use in our calculations.

The angular dependence of the escape probability is exhibited in
Fig. \ref{fig3} for a dust cloud with $\tau_{0.3} = 7$ and three
values of the incident photon wavelength. Note that
$d{\mathcal{P}}^{\rm sc} (\mu)/d\Omega$ remains relatively flat for
$\theta \lsim \theta_{\rm sc}\sim20^{\circ}$ and decreases 
exponentially at larger angles.

\includegraphics{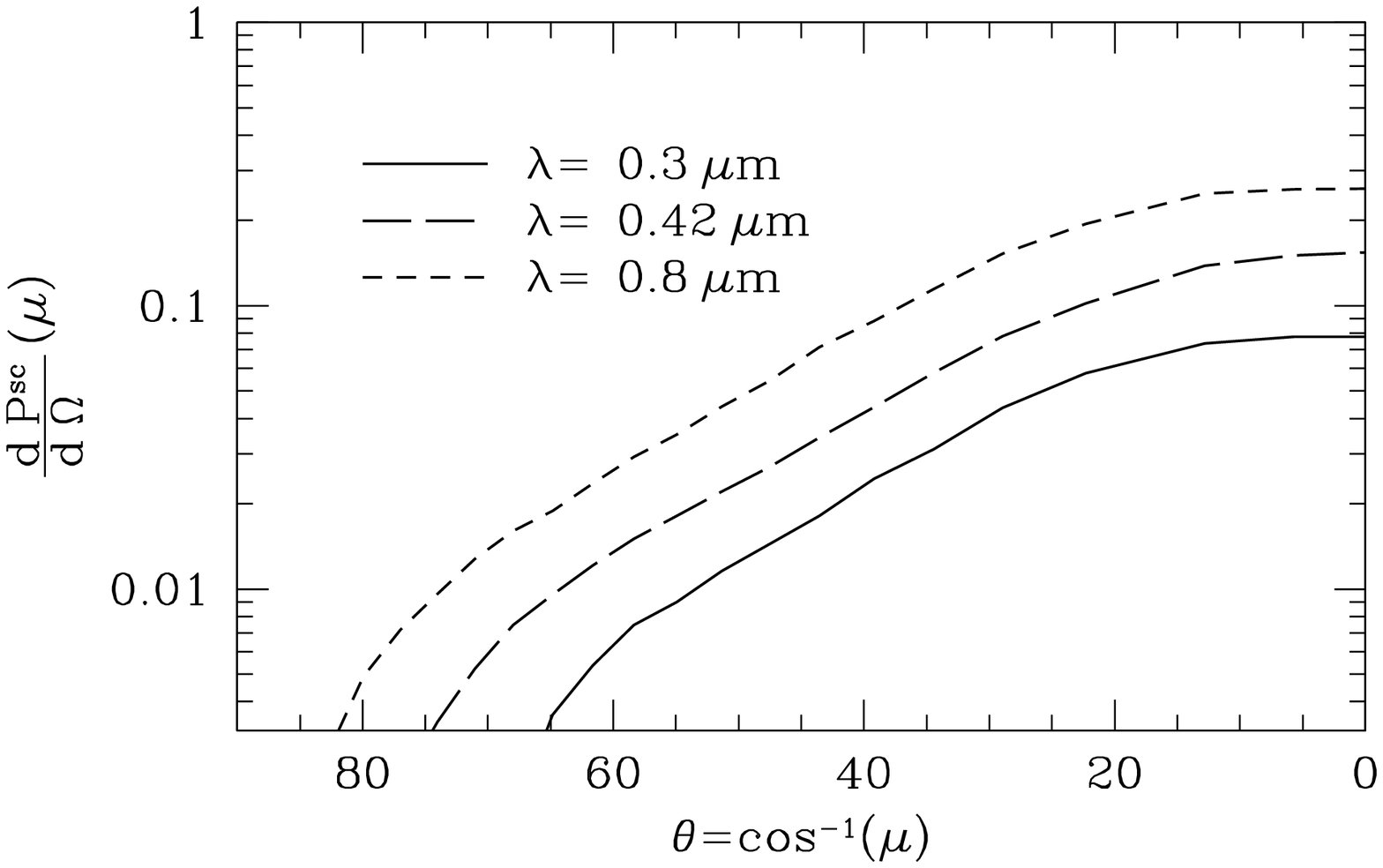} 
\vskip 2.2in 
\figcaption{\label{fig3} The differential escape probability is
plotted as a function of $\theta$ for a dust cloud with
the optical depth for extinction $\tau_{0.3}=7$.  
The results are shown for three different values of the incident
photon wavelength.}
\vskip 0.2in

\subsection{Infrared Echo}
\label{abs}
Hot dust will also emit an isotropic infrared echo due to thermal emission
from dust at the rapid sublimation temperature $\sim2300$~K, peaking
at an observed wavelength $\lambda \sim 2\,(1+z)\,{\rm \mu m}$.
\fcitet{wad99} argue that only the photons in the $1-7.5\,{\rm eV}$
range will contribute to dust heating. For $\tau_{0.3}\sim 7$ the
absorption efficiency for photons in this energy range is $>0.8$; and
moreover, such photons are likely to carry a considerable
fraction of the total OT emission.  Therefore the integrated infrared
flux is
\begin{equation}
\label{FIR}
F^{\rm E}_{\rm IR} = \frac{L\Delta t^{\rm OT}}{8 \pi D_{\rm A}^2} 
\frac{c}{R} \frac{1}{(1+z)^4}.
\end{equation}

\section{Comparison with Observations}
\label{obs}
\subsection{OT-Echo-Redshift Relations}
\label{lum}
Adopting our simple model of dust scattering, Eqs. (\ref{rsubl},
\ref{time}) allow us to relate the sublimation radius and OT power,
$10^{47}L_{47}$~erg s$^{-1}$, to the observed echo delay, $t^{\rm
E}_{\rm ob} \equiv 10^6 t^{\rm E}_{\rm ob\,6}$~s.
\begin{eqnarray}
R_{\rm sub}&\sim&0.2C_1^{-1}C_2^{-1}t^{\rm E}_{\rm ob,6}(1+z)^{-1}\;
{\rm pc},\label{rsub}\\
L_{47}&\sim&0.03(1+z)^{-2}C_1^{-2}C_2^{-2}(t^{\rm E}_{\rm ob,6})^2,
\label{lb47}\;,
\end{eqnarray}
where $C_1=(1-\mu)/0.06$ allows for beaming or
characteristic scattering angles different from $20^\circ$,
and $C_2=R/R_{\rm sub}$ should be used if the dust is located beyond
$R_{\rm sub}$.

For simplicity, we now suppose that the spectral index of the OT is
$\alpha\sim1$. This is quite close to the spectral index of the
observed afterglows.  We can then use Eqs(\ref{Fsc}) to relate the
R-band ($0.65{\rm \mu m}$) echo flux density to the escape probability
\begin{eqnarray}
\label{recho}
F^{\rm E}_\nu[0.65{\rm\mu m}]\sim0.4{dP^{\rm sc}\over d\Omega}
\left({t^{\rm E}_{\rm ob,6}\Delta t^{\rm OT}_{\rm ob,3}
\over C_1C_2^2}\right) \times \\ \nonumber \left({D_{\rm A}
\over1.5{\rm Gpc}}\right)^{-2}(1+z)^{-6}\,{\rm \mu Jy,}
\end{eqnarray}
where the observed duration of the optical transient is $10^3\Delta t^{\rm
OT}_{\rm ob,3}$~s. Note the strong dependence on redshift which
implies that accurate measurements of both the optical transient and
the echo flux could lead to a fairly precise redshift prediction.

The ratios of the optical transient flux density, $F^{\rm OT}_{\nu_{\rm ob}} = 
L_{\nu} f_{\rm ns} (4 \pi)^{-1} D_{\rm A}^{-2} (1+z)^{-3}$, and infrared echo
flux density to the optical echo flux density are likewise given by
\begin{eqnarray}
&&{F^{\rm OT}_\nu[0.65{\rm \mu m}]\over F^{\rm E}_\nu[0.65{\rm\mu m}]}\sim
3000\left({f_{\rm ns}\over C_1dP^{\rm sc}/d\Omega}\right)
\left({t^{\rm E}_{\rm ob,6}\over\Delta t^{\rm OT}_{\rm ob,3}}\right),
\hspace{1cm} \label{burstflux}
\\
&&{F^{\rm E}_\nu[2(1+z){\rm \mu m}]\over F^{\rm E}_\nu[0.65{\rm\mu m}]}\sim
0.5 \left({dP^{\rm sc}\over d\Omega}\right)^{-1}(1+z),
\label{irflux}
\end{eqnarray}
where $f_{\rm ns}$ is the fraction of incident OT photons, emerging 
unscattered from the dust cloud.

\subsection{GRB~980326}
For GRB~980326, an excess R flux $F^{\rm E}_{\nu_{\rm ob}}[0.65{\rm \mu
m}]\sim0.4\,{\rm \mu Jy}$ was measured a time $t^{\rm E}_{\rm
ob}\sim20$~d (\fcitep{bea99}).  If we make the simplest assumptions,
$a_{-1}\sim Q_{\rm abs}\sim C_1\sim C_2\sim1$, then $R_{\rm
sub}\sim0.3(1+z)^{-1}{\rm pc}$ and $L\sim9\times10^{45}(1+z)^{-2}\;
{\rm erg\, s}^{-1}$. Comparing the reported spectral slope,
$\alpha\sim2.8$ of the putative echo to that of the afterglow
($\alpha\sim0.8$), we estimate that $\tau_{0.3}\sim7$ ({\it cf}
Fig. 2). This, in turn, implies that $dP^{sc}/d\Omega$ in the observed
R band $\sim0.2(1+z)^{-1}$ and $f_{\rm ns}\sim0.05(1+z)^{-4}$ (see
Fig. \ref{fig2}). We can then use Eq.(\ref{recho}) to deduce that
$\Delta t^{\rm OT}_{\rm ob,3}\sim3(1+z)^7$ and $F^{\rm OT}_{\nu_{\rm
ob}}[0.65{\rm\mu m}] \sim200(1+z)^{-10}\;{\rm\mu Jy}$. If $z\sim0.4$,
then the energy associated with the first optical measurement of the
afterglow ($F^{\rm OT}_{\nu_{\rm ob}}[0.65{\rm\mu m}] \sim10\;{\rm\mu
Jy}$ after 0.5~d), suffices to account for the observed excess after
20~d as a dust echo. If $z>0.4$, then the optical transient would have
had to be present and create a larger fluence at earlier times. This
is not unreasonable as the OT flux was measured to satisfy $F^{\rm
OT}\propto t^{-2}$. In view of the large number of simplifying
assumptions that we have made, this estimate can only be regarded as
illustrative. However it suffices to demonstrate that dust scattering
is consistent with all of the available data.

\subsection{GRB~970228}
A somewhat similar story can be told for GRB~970228, where the redshift,
$z=0.695$, is known (\fcitep{dea99}). The earliest R-band measurement is 
$\sim30{\rm\mu Jy}$ 0.7~d after the GRB; and after $\sim30$~d there
red excess flux $\sim0.3{\rm\mu Jy}$ was observed, with the spectral slope
($\alpha \sim 3.0$) very similar to that seen in GRB~980326
(\fcitep{gea99}).
For this object, again, within the uncertainties, the fluence measured in the
first stages of the optical transient is sufficient to account for the
energy in the optical excess.
   
\subsection{Dust Origin} 

In both examples above, the mass of dust required to produce an
optical depth $\tau_{0.3}\sim7$ with our simplest assumptions and
assuming that it is spherically symmetrically distributed with respect
to the GRB is $\sim0.1$~M$_\odot$.  This amount of dust could form in
an expanding high-metallicity wind associated with an earlier stage in the
evolution of the GRB progenitor as we have assumed in our simple
model.  Alternatively the dust might be associated with a molecular
cloud if GRBs are associated with massive star formation or a
molecular torus should they be located in obscured galactic nuclei.
\section{Discussion}
\label{impl}

In this letter we present an alternative explanation for the reddened
excess emission observed in GRB~970228 and GRB~980326, which we attribute to
dust scattering of the early-time, afterglow emission.  This scenario
is predictive enough to be confirmed or ruled out with observations of
future GRBs. In particular, in contrast to the supernova explanation
(\fcitep{bea99}; \fcitep{gea99}; \fcitep{rei99}), if the excess
emission is due to dust scattering, then its properties will depend on
the luminosity of the optical transient.  HETE II
(http://space.mit.edu/HETE/) scheduled to be launched in early 2000
and Swift (http://swift.gsfc.nasa.gov/homepage.html), scheduled for
2003, should provide real-time localization of GRB X-ray afterglows
with sufficient precision to permit faster follow-up and better
measurements of its total fluence.  Infrared observations may discover
the expected thermal emission from hot subliming dust ({\it cf}
\fcitep{wad99}).  In
fact dust emission might be the correct explanation for the
``near-IR'' bump seen in the spectrum of the GRB~991216 afterglow
(\fcitep{fea00}).  Note that as most GRBs are at redshifts $\gsim 0.5$,
3~${\rm\mu m}$, (as as opposed to the more common 2~${\rm\mu m}$)
photometry may be necessary to see this emission.

In those GRBs, where it is also possible to measure a redshift, the
the simplest model of dust-scattering is over-contrained and therefore
refutable. Beaming and dust inhomogeneity introduce additional
uncertainty but such models may also be excludable. For example, if
ROTSE (\fcitep{aea99}) were to detect another optical flash in a GRB as
luminous as that seen in GRB~990123, which had an isotropic luminosity
$L\sim10^{51}$~erg s$^{-1}$, then dust should be physically sublimed
out to a distance $R_{\rm sub}\sim100$~pc along the line of sight.
Unreasonably large beaming would then be required to explain a dust
echo with a delay of only a few weeks. Alternatively, if the radio
light curve in an afterglow tracked the optical light curve, then this
would be incompatible with both dust scattering and a supernova.

A further prediction of the dust echo model is that, unless the dust
and OT are both arranged axisymmetrically with respect to the line of
sight, we expect there to be linear polarization associated with dust
echos and this may be measurable in bright examples.  (1.7 percent
polarization has been reported in the optical transient associated
with GRB 990510 by \fcitet{cov99} but this is unlikely to be due to
scattering.)

In conclusion, we have demonstrated that dust scattering can account for the
excess optical emission observed in the afterglows 
of two GRBs as an alternative to an underlying supernova explosion.
Future observations should be able to rule out or confirm this explanation.
\acknowledgements We thank J. Bloom for helpful comments.
This work was supported by NASA grant 5-2837, NSF grant 
AST 99-00866 and a
Chandra Postdoctoral Fellowship grant \#PF8-10002 awarded by the
Chandra X-Ray Center, which is operated by the SAO for NASA under
contract NAS8-39073.

\bibliography{grb}

\vfill\eject

\end{document}